\documentclass[aps,prb,twocolumn,superscriptaddress,longbibliography]{revtex4-1}
\usepackage{epsfig,amsopn}
\usepackage{graphicx}
\usepackage{color}
\usepackage{amsmath,amssymb}
\usepackage{enumerate}
\newcommand\bea{\begin{eqnarray}}
\newcommand\eea{\end{eqnarray}}
\newcommand\beq{\begin{equation}}
\newcommand\eeq{\end{equation}}

\def\nn{\nonumber}
\def\f{\frac}
\def\al{\alpha}

\def\ep{\epsilon}

\def\si{\sigma}
\def\Do{\partial}

\def\ra{\rangle}
\def\ua{\uparrow}
\def\da{\downarrow}

\def\th{\theta}

\begin{document}
\title{Finite transverse conductance and anisotropic magnetoconductance under an applied in-plane magnetic field in two-dimensional electron gases with strong spin-orbit coupling} 
\author{ Abhiram Soori~}
\email{abhirams@uohyd.ac.in}
\affiliation{School of Physics, University of Hyderabad, C. R. Rao Road, Gachibowli, Hyderabad-500046, India}
\begin{abstract}
The current in response to a bias in certain two-dimensional electron gas (2DEG),  can have a nonzero transverse component under a finite magnetic field applied in the plane where electrons are confined. This phenomenon known as planar Hall effect is accompanied by dependencies of both the longitudinal and the transverse components of the current on the angle $\phi$ between the bias direction and the magnetic field. This effect can be observed in a variety of systems, for example in topological insulators where spin-momentum locking of the topologically protected surface states is the root cause for the effect and in magnetic systems where anisotropic magnetic ordering  induces it. In  2DEG with spin orbit coupling (SOC) such as oxide interfaces, this effect has been experimentally witnessed. Further, a fourfold oscillation in longitudinal resistance as a function of $\phi$ has also been observed. Motivated by these, we perform scattering theory calculations on a 2DEG with SOC in presence of an in-plane magnetic field connected to two dimensional leads on either sides to obtain longitudinal and transverse conductances. We find that the longitudinal conductance is $\pi$-periodic and the transverse conductance is $2\pi$-periodic in $\phi$. The magnitude of oscillation in transverse conductance with $\phi$ is enhanced in certain patches in $(\alpha,b)$-plane where $\al$ is the strength of SOC and $b$ is Zeeman energy due to magnetic field. The oscillation in transverse conductance with $\phi$ can be highly multi-fold for large values of $\alpha$ and $b$. The highly multi-fold oscillations of transverse conductance are due to Fabry-P\'erot type interference between the modes in the central region as backed by its length dependent features. Our study establishes that SOC in a material is sufficient to observe planar Hall effect without the need for anisotropic magnetic ordering or nontrivial topology of the bandstructure. 
\end{abstract}
\maketitle
\section{Introduction}
The longitudinal and the Hall resistances in Hall measurements of certain systems in presence of a magnetic field applied in the same plane as Hall measurements depend on the angle between  magnetic field and  the longitudinal direction. Such a dependence of the longitudinal resistance/conductance is called anisotropic magneto-resistance/conductance, whereas such a dependence of the Hall resistance is called planar Hall effect.  The  anisotropic magnetoresistance~(AMR) and planar Hall effect~(PHE) have been observed in a variety of magnetic systems~\cite{nazmul08,Li10,Roy10,annadi13}. Recently PHE has also been observed in magnetic skyrmion systems~\cite{hirschberger20}. PHE amplitude in certain magnetic systems is sensitive to magnetic fields as low as earth's magnetic field~\cite{Roy10}. Recently, experimental investigations of AMR and PHE have been in exotic topological materials such as topological insulators~\cite{taskin17,rakhmi18,he19,bharadwaj21} and Weyl semimetals~\cite{burkov17,kumar18}. A common physical factor among topological materials is spin momemntum locking which has origins in strong spin orbit coupling~(SOC) of the constituent atoms~\cite{qi11,armitage18}. The SOC in the bulk of topological insulators such as Bi$_2$Se$_3$ is Dresselhaus like which results in  topologically protected surface states that pocess spin-momentum locking~\cite{qi11}. The spin-momentum locking is responsible for PHE and AMR in topological insulators~\cite{suri21}. A natural question that arises is: `does a system with SOC alone, but topologically trivial exhibit PHE and AMR?' It is therefore interesting to investigate AMR and PHE in two-dimensional electron systems with SOC. 

A two dimensional electron gas~(2DEG) with SOC is an ideal platform  for the realization of Datta-Das transistor proposed in 1990~\cite{dattadas}. But such systems are not easy to achieve experimentally since existence of Rashba spin split bands alone with large enough SOC strength is rare. In 2015, Datta-Das transistor was realized experimentally in InGaAs/InAlAs heterostructures~\cite{chuang2015}. Recently, $\mathrm{Bi_2 Se_3/MoTe_2}$ heterostructure has been proposed to be a 2DEG with large SOC strength~\cite{wang2017}. Another 2DEG with large SOC strength is LaAlO$_3/$SrTiO$_3$ interface where the strength of SOC can be further controlled by an applied gate voltage~\cite{annadi13}. In addition to SOC, there are localized magnetic impurities that determine the transport behavior in this system.  This system exhibits PHE and a peculiar type of AMR. The AMR oscillation is fourfold when the electrons confined at the interface is purely two-dimensional whereas it is twofold when the interfacial electrons have access to three dimensions. Also, AMR has been extensively studied in LaAlO$_3/$SrTiO$_3$ interface~\cite{rout17} and in SrTiO$_3$~\cite{miao16}.  PHE and AMR have been observed recently at the interface of LaVO$_3$-KTaO$_3$ which hosts electrons confined to two dimensions with large SOC~\cite{wadehra20}. In this system, the AMR oscillation is twofold at low magnetic fields whereas it is fourfold at high magnetic fields. AMR has been investigated in 2DEG with SOC in a number of studies~\cite{wang09,wang10,wang11,boudjada19}, taking into account scattering from impurities. But a theoretical study of PHE in 2DEG with SOC is missing. 

\begin{figure}
 \includegraphics[width=6.5cm]{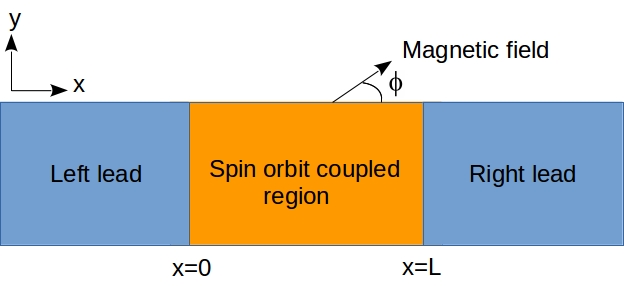}
 \caption{Schematic diagram of the setup under study. The region in the middle from $x=0$ to $x=L$ is the region with spin-orbit coupling and an in-plane magnetic field which makes an angle $\phi$ with the longitudinal direction. A bias is applied from left lead to right lead. }\label{fig:schem}
\end{figure}

In this work, we study transport across a 2DEG with SOC in presence of a magnetic field applied in the same plane as 2DEG,  connected to two-dimensional leads on either sides as shown in fig.~\ref{fig:schem}. We follow the scattering theory approach developed in 
a recent work in the context of topological insulators~\cite{suri21} to study AMR and PHE quantitatively by calculating  the longitudinal and transverse conductances. The difference between this approach and  the experiments is that here the components of the current in the longitudinal and transverse directions in response to the bias in longitudinal direction are determined whereas in experiments, voltages developed in longitudinal and transverse directions in response to current in longitudinal direction are measured. In this paper, we present the details of calculation in section~\ref{sec:calc} and present the results accompanied by an analysis in section~\ref{sec:result}. In section~\ref{sec:concl}, we discuss the main results and conclude.  

\section{Details of calculation}~\label{sec:calc}

The Hamiltonian for the system under study is
\bea H &=& \Big[-\f{\hbar^2}{2m}\Big(\f{\Do^2}{\Do x^2}+\f{\Do^2}{\Do y^2}\Big)-\mu\Big]\si_0 +\Delta_{L}(x)\Big[i\al\Big(\si_y\f{\Do}{\Do x}\nn \\ &&-\si_x\f{\Do}{\Do y}\Big)+b(\cos\phi\si_x+\sin\phi\si_y)\Big], \label{eq:ham}\eea
where $\Delta(x)=\Theta(x)\Theta(L-x)$ and $\Theta$ is Heavyside step function. Here, $m$ is the effective mass of electrons in the system, $\mu$ -the chemical potential, $\al$ -the strength of SOC, $b$ -the magnitude of the Zeeman energy due to the magnetic field,  $\phi$ -the angle made by the in-plane magnetic field with the $x$-axis and $\si_x,\si_y$ are Pauli spin matrices. The SOC and the magnetic field are present only in the region $0<x<L$, and the regions to the left and right model the leads. The length of the system in $y$-direction is $L_y$ and the limit $L_y\to\infty$ is taken along with periodic boundary condition in $y$-direction. We do not assume any barrier at the interfaces $x=0,L$. The boundary conditions at $x=0,L$ are different from the conventional continuities of the wavefunction $\psi$ and its derivative $\Do_x\psi$. These boundary conditions can be derived by demanding the conservation of the $x$-component of current at the junctions $x_0=0,L$. One choice of the boundary conditions is the continuity of the wavefunction $\psi$ accompanied by  
\bea \Do_x\psi|_{x_0^{-s_0}} &=& \Do_x\psi|_{x_0^{s_0}} -\f{i\al m}{\hbar^2} \si_y\psi|_{x_0^{s_0}}, \label{eq:BC} \eea
where $s_0={\rm sign}(L/2-x_0)$ and $x_0^{s_0}=lim_{\ep\to 0^+}[x_0+s_0\ep]$. The dispersion in the regions $x<0$ and $x>L$ are $E=\hbar^2(k_x^2+k_y^2)/2m-\mu$, while the dispersion in the central region ($0<x<L$) is: $E=\hbar^2(k_x^2+k_y^2)/2m-\mu\pm\sqrt{(b\cos{\phi}+\al k_y)^2+(b\sin{\phi}-\al k_x)^2}$. The wavefunction at energy $E$ of an incident $\si$-spin electron from the left lead at an angle $\th$ with $x$-axis has the form  $e^{ik_yy}\psi_{\si}(x)$, where  
\bea \psi_{\si}(x) &=& e^{ik_xx}|\si\ra + \sum_{\si'=\ua,\da} r_{E,\si'\si}~e^{-ik_xx}|\si'\ra, {\rm ~for~}x<0, \nn \\ &=&\sum_{j=1}^4s_{E,j,\si}e^{ik'_{xj}x}[u_j,v_j]^T, ~~{\rm for~} 0<x<L,  \nn \\ &=& \sum_{\si'=\ua,\da} t_{E,\si'\si}~e^{ik_xx}|\si'\ra, ~~{\rm for~} x>L, \eea 
$k_y=\sqrt{2m\mu}\sin{\th}/\hbar$, $k_x=\sqrt{2m\mu}\cos{\th}/\hbar$, $\si'$ is  the spin opposite to $\si$, $|\ua\ra=[1,0]^T$, $|\da\ra=[0,1]^T$ and $k'_{xj}$ (for $j=1,2,3,4$) are four roots of the dispersion of the central region for $k_x$ at the given $E$ and $k_y$. Due to translational invariance in $y$-direction, $k_y$ is same in all the regions. The current component along $x$-direction is same everywhere whereas the current component along $y$-direction can vary as a function of spatial location $x$. In the region $0<x<L$, at location $x$, the $y$-component of current due to an incident $\si$-spin electron at energy $E$ that makes an angle $\th$ with $x$-axis is: \bea I_{y,\si}(E,\th,x) &=& e\Big[\f{\hbar k_y}{m}\psi_{\si}^{\dag}(x)\psi_{\si}(x) +\f{\al}{\hbar}\psi_{\si}^{\dag}(x)\si_x\psi_{\si}(x)\Big],~~~   \label{eq:Iy} \eea
 where $e$ is the electron charge. Let $[I_x,I_y(x)]$ be the current flowing at $x$ in response to a voltage bias in applied in the window $(0,eV)$. Then, the differential- longitudinal and  transverse conductances are $G_{xx}=dI_x/dV$ and $G_{yx}(x)=dI_y(x)/dV$ which can be expressed as
\bea G_{xx}~~ &=& \f{\sqrt{2m(\mu_N+eV)}}{\hbar}\f{e^2}{h}L_y\sum_{\si,\si'}\int_{-\pi/2}^{\pi/2}d\th\cos{\th} |t_{\si',\si}|^2, \nn \\ G_{yx}(x) &=& \f{e}{h}\f{mL_y}{\hbar}\sum_{\si}\int_{-\pi/2}^{\pi/2}d\th~I_{y,\si}(eV,\th,x), {\rm ~for~}0<x<L, \nn \\ G_{yx}(x) &=& \f{e^2}{h}L_y\sum_{\si,\si'}\int_{-\pi/2}^{\pi/2}d\th~ k_y|t_{\si',\si}|^2 , ~~{\rm for~}x>L.  \label{eq:Gxx-Gyx} \eea
The dependence of $G_{xx}$ on $\phi$ is termed anisotropic magnetoconductance~(AMC) and the dependence of $G_{yx}$ on $\phi$ is termed PHE. PHE amplitude is defined as half the difference between maximum and minimum values of $G_{yx}$ when $\phi$ is varied. 

\begin{figure*}
 \includegraphics[width=5.8cm]{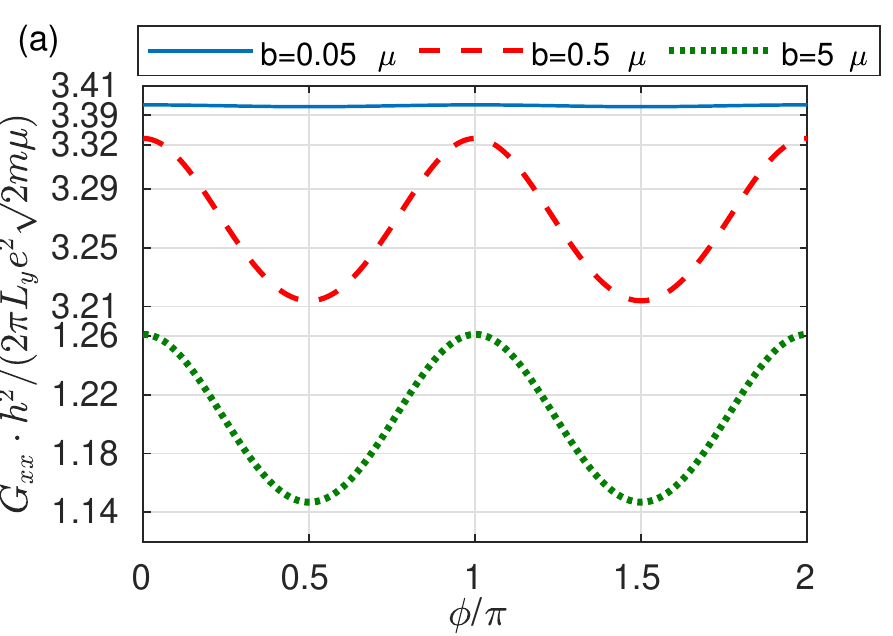}
   \includegraphics[width=5.8cm]{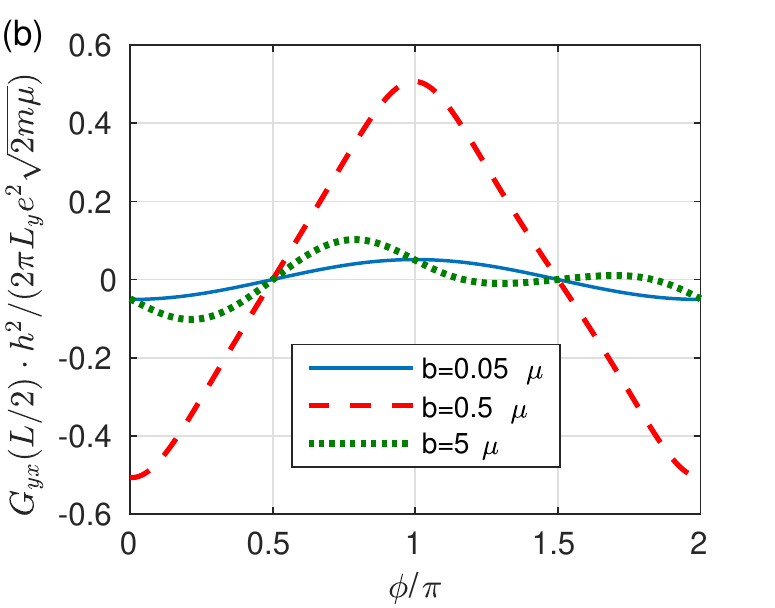}
 \includegraphics[width=5.8cm]{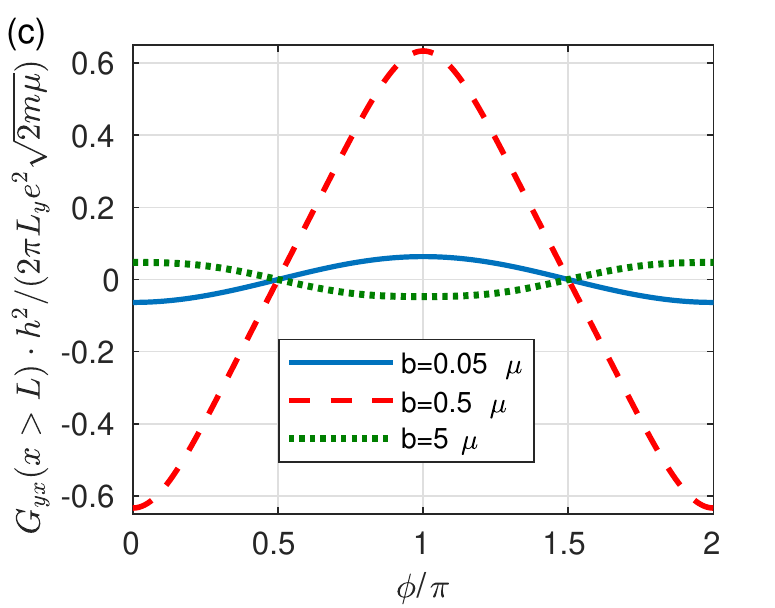}
 \includegraphics[width=5.8cm]{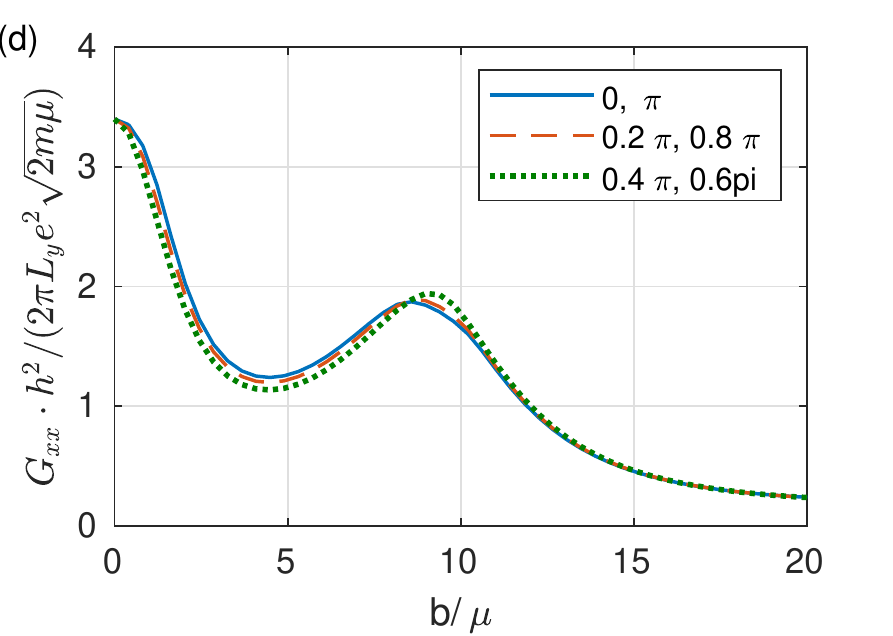}
 \includegraphics[width=5.8cm]{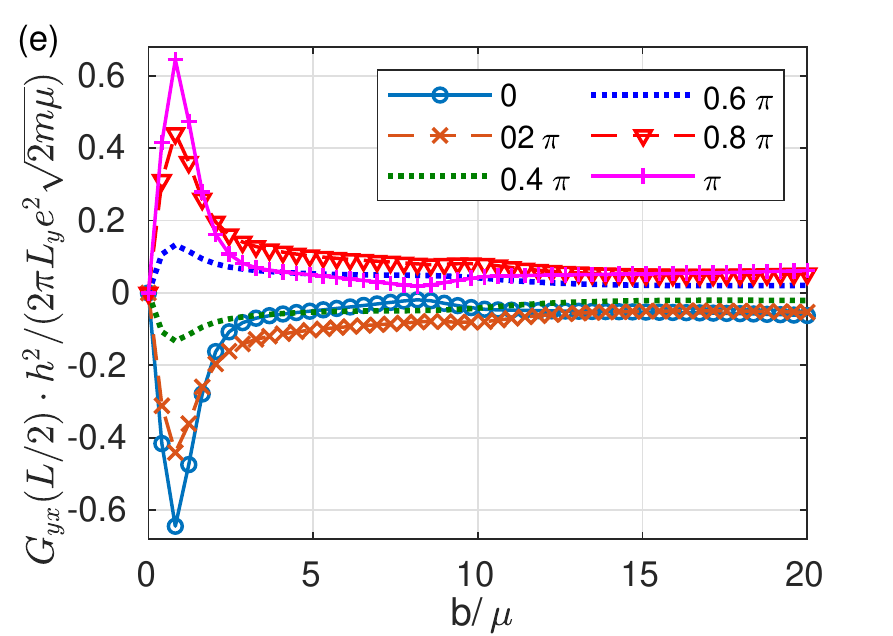}
 \includegraphics[width=5.8cm]{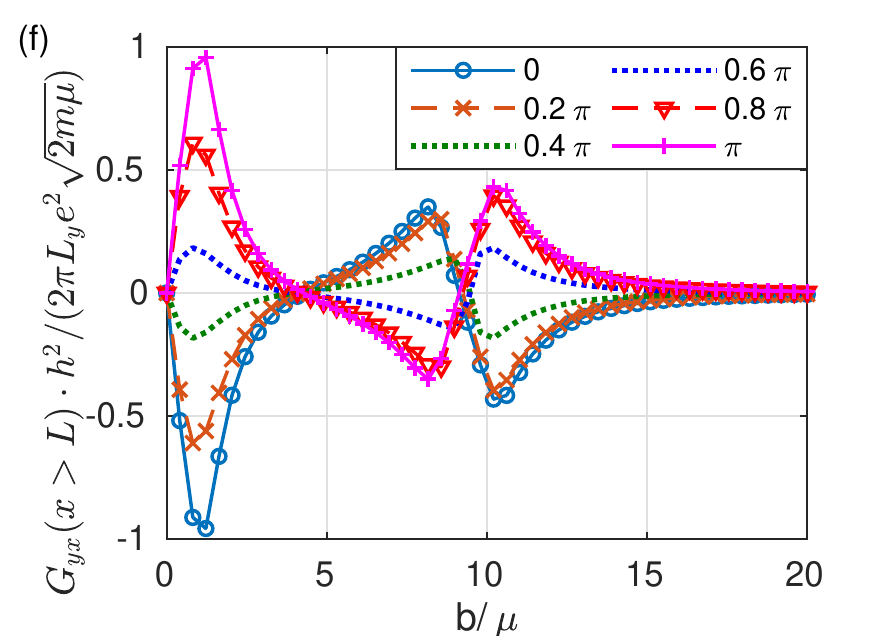}
\caption{Anisotropic magnetoconductance and planar Hall effect. 
(a) The longitudinal conductance, and the transverse conductances:
at (b) $x=L/2$ and in the region (c) $x>L$ as plotted as functions
of $\phi$ - the angle between applied in-plane magnetic field and 
the longitudinal direction. In (a), different curves are displaced
along $y$ so as to fit all curves in the same graph and show the 
dependence of $G_{xx}$ with $\phi$. In doing so, the values on 
$y$-axis are also relabeled - the gaps in ranges of $y$ axis 
$(1.26,3.21)$ and $(3.32,3.39)$ are not shown. In (d) longitudinal conductance, in (e) transverse conductance at $x=L/2$ and in (f) transverse conductance at $x>L$ are plotted as functions of  $b$ for different choices of angles $\phi$ mentioned in the plot legend.   Parameters: $\al=\hbar\sqrt{\mu/(2m)}$, $L=\hbar/\sqrt{2m\mu}$.}~\label{fig:gvsphib}
\end{figure*}

\section{Results and Analysis}~\label{sec:result}

\begin{figure}
 \includegraphics[width=7.1cm]{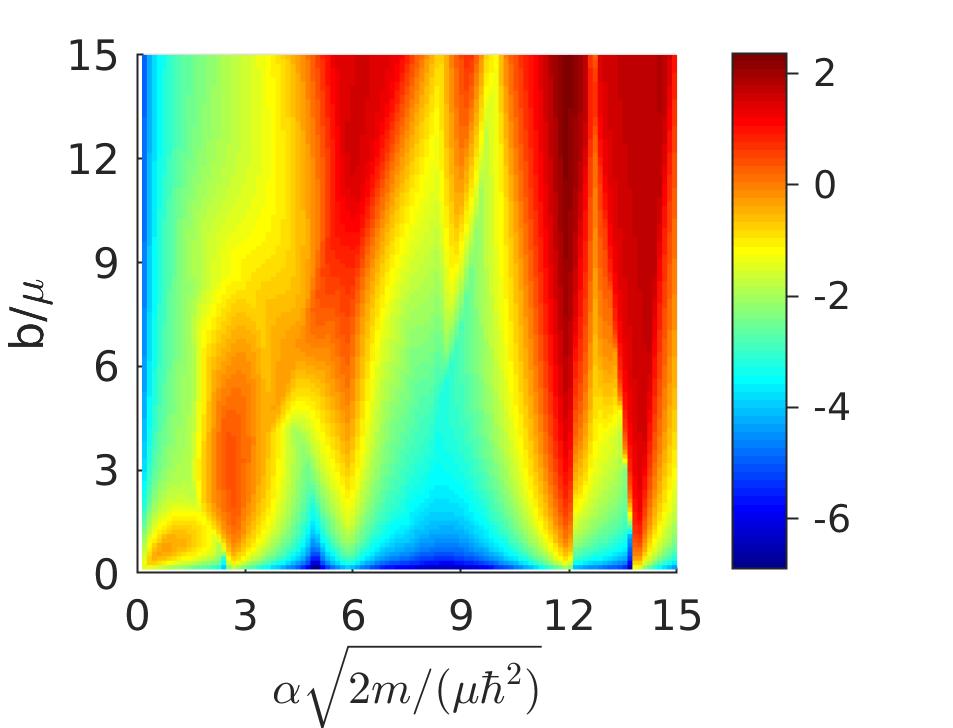}
 \caption{Logarithm of PHE amplitude evaluated at $x=L/2$ as a function of $b$ and $\al$. $L=\hbar/\sqrt{2m\mu}$.}~\label{fig:phea}
\end{figure}
Parameters in the Hamiltonian eq.~\eqref{eq:ham} are $m$, $\mu$, $L$, $\al$ and $b$. All parameters apart from $m$ and $\mu$ are expressed in terms of these two parameters. $L$ is taken to be $\hbar\sqrt{2m\mu}$, except when mentioned specifically. Since current along $x$-axis is conserved, the longitudinal conductance $G_{xx}$ is same at all locations $x$. But the transverse conductance $G_{yx}$ is a function of the spatial location $x$. All the conductances calculated in this work are at zero bias (i.e., at $E=0$). 
In Fig.~\ref{fig:gvsphib}(a,b,c), we plot $G_{xx}$, $G_{yx}(L/2)$ and $G_{yx}(x>L)$, all at zero bias as functions of $\phi$ respectively at different strengths  of the magnetic field labeled in the legend.  From fig.~\ref{fig:gvsphib}(a), we can see that for low magnetic fields, the amplitude of variation in AMC is small and it increases with magnetic field accompanied by a  decrease in the mean value of AMC. The mean value of AMC decreases with magnetic field  since a higher magnetic field implies a higher mismatch between the wavenumbers in different regions. In fig.~\ref{fig:gvsphib}(b,c) the transverse conductance shows oscillations of a higher magnitude for an intermediate magnetic field. At low magnetic fields, the transverse drift of the incident current is small whereas at large magnitudes of the magnetic field $b$, the SOC becomes relatively small leading to a lesser importance of spin being locked to the momentum. For $b=5\mu$, we see that the transverse conductance at $x=L/2$ exhibits two-fold oscillation, whereas it is still one-fold at $x>L$. The sign of $G_{yx}(x>L)$ at $\phi=0$ changes with magnetic field strength. For angles $\phi=\pi/2,3\pi/2$, it can be seen from the system Hamiltonian that the system becomes symmetric under $y\to-y$. Hence, the transverse conductance at these angles is exactly zero as can be seen in fig.~\ref{fig:gvsphib}(b,c). In fig.~\ref{fig:gvsphib}(d,e,f), we plot three conductances $G_{xx}$, $G_{yx}(x=L/2)$ and $G_{yx}(x>L)$ respectively as functions of  $b$. The overall decrease in $G_{xx}$ with increasing magnetic field is due to the increasing mismatch between the wavenumbers in different regions. But the local peak around $b\sim 9\mu$ is due to the Fabry-P\'erot interference in the central region, which we have verified by changing the length $L$ in the system. We see that the magnitude of transverse conductance at angles $\phi$ apart from $\pi/2, 3\pi/2$ first increases with magnetic field, reaches a peak and then decreases. We see that particularly the $G_{yx}(x>L)$ oscillates with $b$ changing sign at certain values of $b$ with an overall decrease in local peak. Such oscillations in $G_{yx}(x>L)$ with $b$ is are due to Fabry-P\'erot type interference in the central region as can be verified by varying the length $L$~\cite{soori12,soori17,nehra19,soori19}.

\begin{figure}
 \includegraphics[width=7cm]{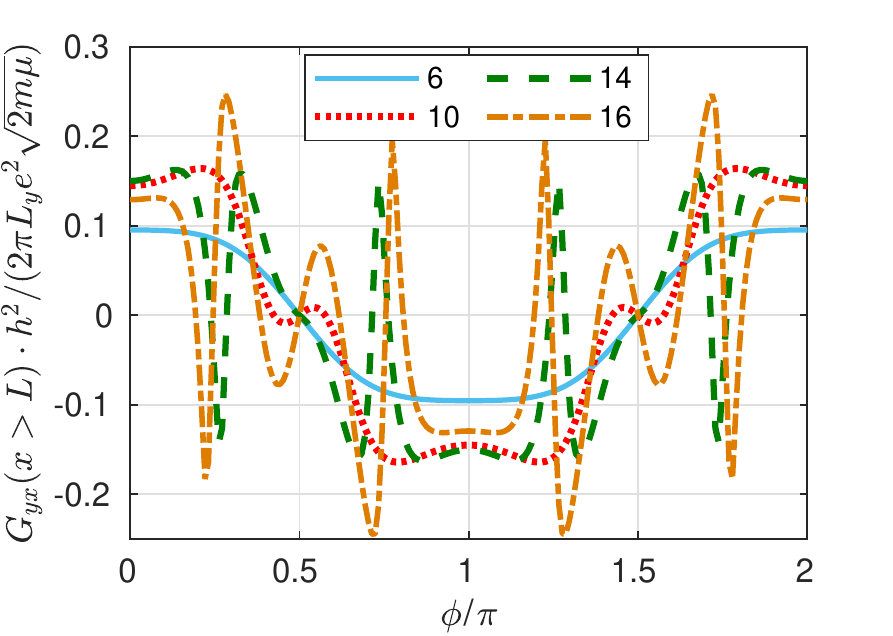}
 \caption{$G_{yx}(x>L)$ plotted as a function of $\phi$ for the choice $\al=9\hbar\sqrt{\mu/(2m)}$ and for different values of $b/\mu$ mentioned in the legend. $L=\hbar/\sqrt{2m\mu}$.}~\label{fig:gyxr_al_9}
\end{figure}

\begin{figure}
 \includegraphics[width=7cm]{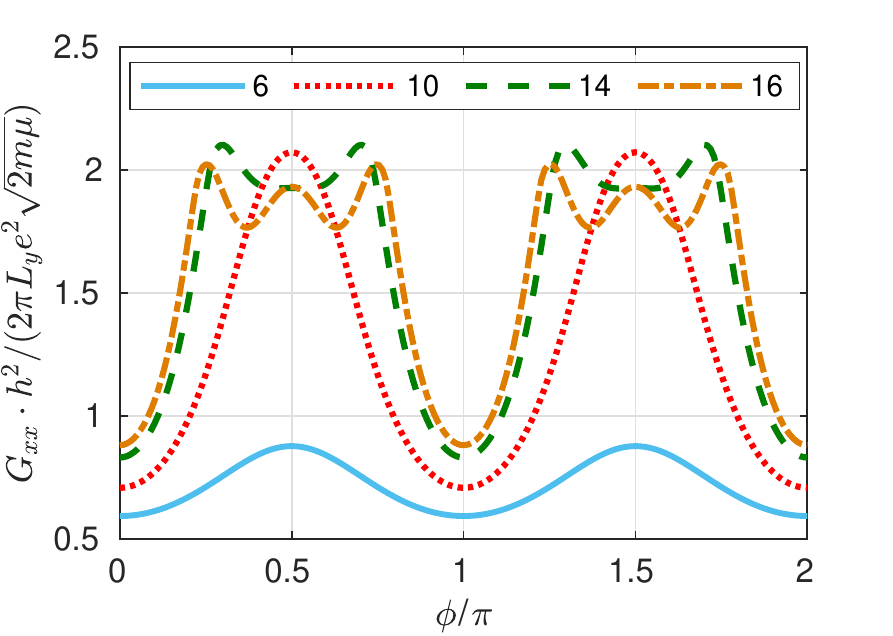}
 \caption{$G_{xx}$ plotted as a function of $\phi$ for different choices of $b/\mu$ shown in the legend for the same choice of parameters as in fig.~\ref{fig:gyxr_al_9}.} \label{fig:gxx_al_9}
\end{figure}

In Fig.~\ref{fig:phea}, we show a color plot of the logarithm of PHE amplitude evaluated at $x=L/2$ as a function $b$ and $\al$, keeping the other parameters the same. We see multiple patches in $(\al,b)$-plane where PHE amplitude is high. The transverse conductance at $x>L$ shows interesting features in these regions. In the patch between $0<\al\sqrt{2m/(\mu\hbar^2)}\lesssim 2$, $G_{yx}(x>L)$ shows one-fold oscillation similar to the one in fig.~\ref{fig:gvsphib}(c), whereas $G_{xx}$ shows two-fold oscillations. In the high PHE amplitude patch in the range $2\lesssim \al\sqrt{2m/(\mu\hbar^2)} \lesssim4$, the oscillation of $G_{yx}(x>L)$ with $\phi$ is one-fold at low $b$ ($b\le\mu$) and becomes three-fold in the middle of the patch as $b$ lies in the range $(2,6)\mu$. At higher values of $b$ in this range of $\al$, $G_{yx}(x>L)$ is one-fold. Here, we use the term $n$-fold for the oscillation of $G_{yx}(x>L)$ with $\phi$ if the value of $G_{yx}(x>L)$ crosses the mean value of $G_{yx}(x>L)$ (which is $0$) $2n$-times as $\phi$ is varied across the range $[0,2\pi]$. We now explore the narrow patch around $\al\sqrt{2m/(\mu\hbar^2)}\sim 9$ in fig.~\ref{fig:gyxr_al_9}. We can see that as $b/\mu$ increases from $6$ to $16$, the oscillation in $G_{yx}(x>L)$ changes from being one-fold to seven-fold. For the choice of $b/\mu=6,10,14,16$ the oscillation in $G_{yx}(x>L)$ is one-, three-, five-, seven-fold respectively. 
For the same set of parameters, the longitudinal conductance $G_{xx}$ shows a number of local peaks as a function of $\phi$ that increases with $b$ as can be seen in fig.~\ref{fig:gxx_al_9}. For choices of $b/\mu=6,10,14,16$, $G_{xx}$ shows two, two, four, six local peaks respectively. This is a reflection of highly multi-fold oscillation in $G_{yx}(x>L)$ though not an exact one-to-one correspondence. 

\begin{figure}
 \includegraphics[width=8cm]{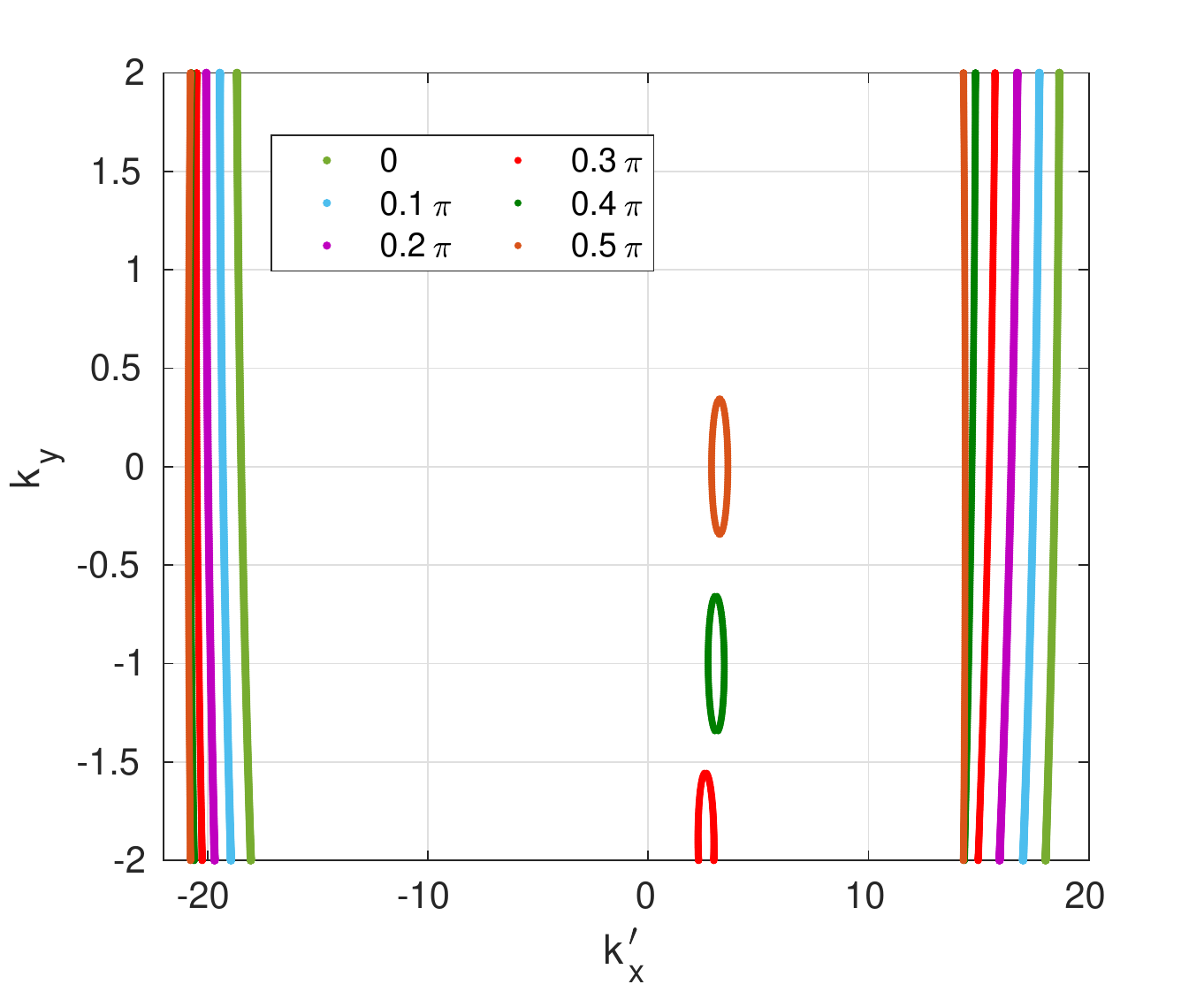}
 \caption{Part of the Fermi surface plotted across a range of $k_y$ that are allowed in the leads. Different curves are for different choices of $\phi$ as shown in the legend. $\al=9\hbar\sqrt{\mu/(2m)}$ and $b=9\mu$. }~\label{fig:fermiarc}
\end{figure}

It is instructive to look at the evolution of the Fermi-surface of the central region as the angle $\phi$ between longitudinal direction and the magnetic field is varied. For the choice of parameters $\al=9\hbar\sqrt{\mu/(2m)}$ and $b=14\mu$, we plot the Fermi surface for different choices of $\phi$ in the range $[0,0.5\pi]$ in fig.~\ref{fig:fermiarc}, focusing on the range of $k_y$ allowed by the leads. Outside this range of $k_y$, there is no scattering.  Because of spin degree of freedom, there are two bands and hence we expect two Fermi surfaces. For small values of $\phi$, there is only one Fermi surface in the focused range of $k_y$ and the second (inner) Fermi surface  appears in this range of $k_y$ around $\phi\lesssim0.3\pi$. For the choice $\phi=0.5\pi$, the Fermi surfaces are symmetric about the line $k_y=0$ and hence the transverse conductance in zero for this choice of $\phi$. For other choices of $\phi$, there exists asymmetry between modes at $k_y$ and $-k_y$ as can be seen from this figure, and this is the root cause for finite transverse conductance.

\begin{figure}
 \includegraphics[width=8cm]{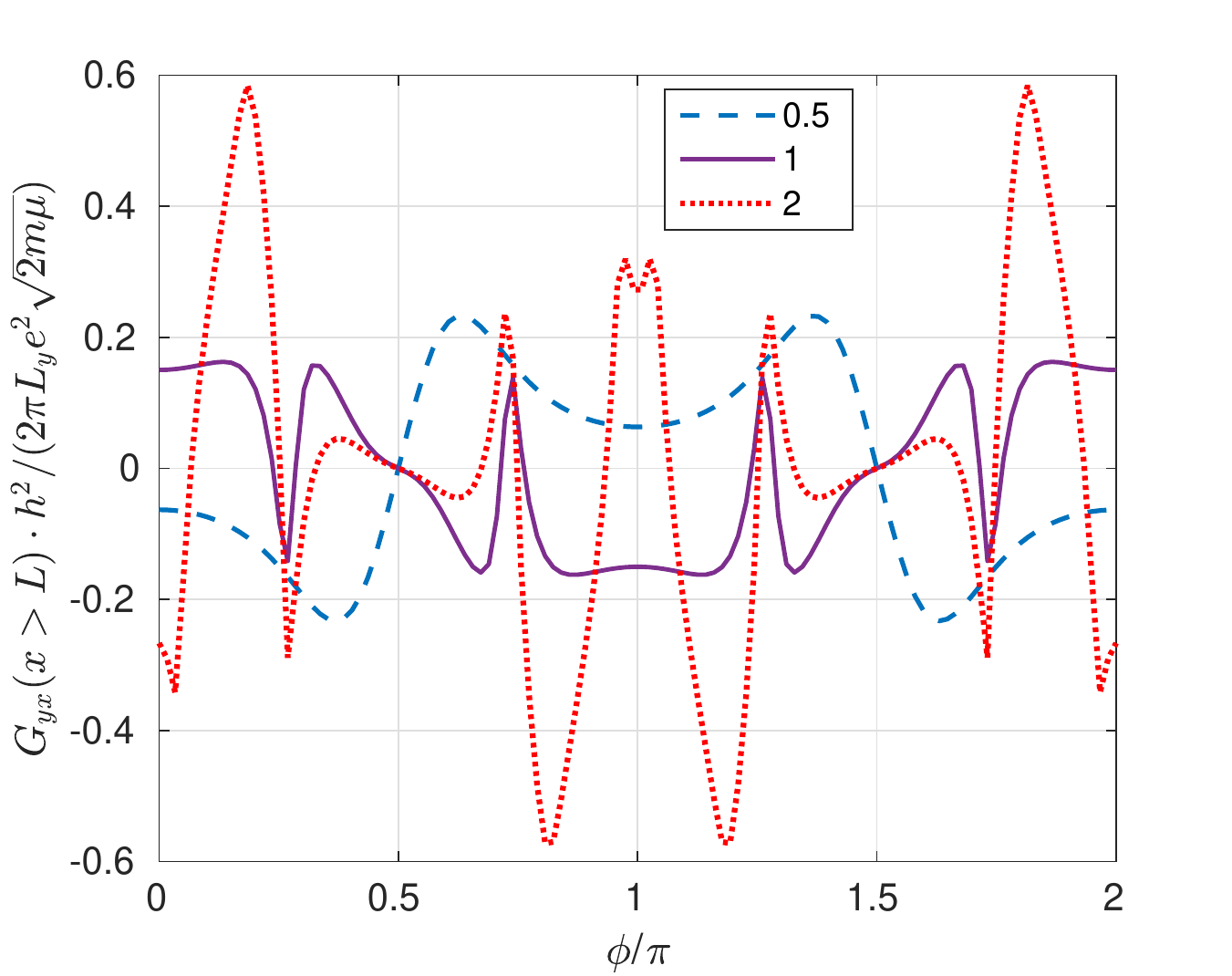}
 \caption{$G_{yx}(x>L)$ versus $\phi$ for different lengths $L$ of the SOC region in the middle for $\al=9\hbar\sqrt{\mu/(2m)}$ and $b=9\mu$. Different curves are for different lengths of the central region indicated by the values of $L\sqrt{2m\mu}/\hbar$ in the plot legend.} \label{fig:gyxr-L}
\end{figure}

Now we turn to the origin of high-fold oscillations in $G_{yx}(x>L)$ at large strengths of SOC and high magnetic fields. For a given $\phi$ and a given angle  $\th$ of incident electron, $k_y$ is fixed and the transport problem at the heart of the phenomenon is that of transmission across a region of SOC and magnetic field in effectively one dimension. Hence, the transmission probability is determined by the interference condition which is more complex than the simpler condition for transmission across a one dimensional barrier $k''_xL=n\pi$, where $k''_x$ is the wavenumber in the barrier region. Further, the value of transverse conductance in the region $x>L$ is determined by adding $k_y$ times the transmission probabilities over different angles of incidence $\th$. So, effectively we can write a Fabry-P\'erot interference condition $k''_x(\phi,L)L=n\pi$, where $k''_x(\phi,L)$ depends on $\phi$ and $L$. For larger values of $b$ and $\al$, the variation of $k''_x$ with $\phi$ is more drastic and hence a higher-fold oscillation in $G_{yx}(x>L)$. The dependence of $k''_x$ on $L$ is slow beyond a certain length since the transverse conductance is obtained by integrating over all angles of incidence. As $L$ increases, $G_{yx}(x>L)$ is expected to show higher-fold oscillations as a function of $\phi$. In fig.~\ref{fig:gyxr-L}, we see higher-fold oscillation in $G_{yx}(x>L)$ as $L$ is increased in qualitative agreement with the above argument. 

\section{Discussion and Conclusion}\label{sec:concl}
We have studied the phenomena of PHE and AMC in 2DEG with SOC.  In magnetic materials, PHE occurs as a result of anisotropic magnetic texture~\cite{nazmul08,Li10,Roy10,hirschberger20}. PHE observed in Weyl semimetals~\cite{burkov17,kumar18,nandy17} is attributed to the chiral anomaly which has origin in topology of the bandstructure. In topological insulators, PHE is ascribed to contain the information about the topological surface state transport~\cite{taskin17}. In topological insulators, PHE and AMR result from perfect spin-momentum locking in the topologically protected surface states~\cite{suri21}. In contrast, we have shown in this work that PHE and AMR can be observed in systems with SOC alone without the need for topological protection or magnetism. 
Further, we have shown that PHE can have a large magnitude at large values of  SOC strength and magnetic field. At a given strength of SOC, at very small and very large  magnetic fields, the PHE amplitude is small. When the PHE amplitude evaluated at $x=L/2$ is large, the oscillations in transverse conductance evaluated in the region $x>L$ can be highly multi-fold. We find one-, three-, five-, and seven-fold oscillations of the transverse conductance. This is in contrast to the fourfold oscillation of the AMR in systems with SOC~\cite{annadi13,wadehra20}. In the first place, PHE is due to breaking of the $y\to-y$ symmetry in the Hamiltonian when an in-plane magnetic field is applied in the region with SOC. Further, the highly multi-fold oscillations in transverse conductance is due to the Fabry-P\'erot type interference in the central region. Such oscillations also show up in AMC, but to a lower degree. Our results do not apply to LaAlO$_3/$SrTiO$_3$ interface despite strong SOC since magnetic impurities are present in this system.  PHE can be employed in detection of magnetic fields making the highly multi-fold oscillations useful in development of magnetic field sensors. PHE has not been widely observed in two-dimensional electron systems with SOC due to small strength of SOC. LaVO$_3$-KTaO$_3$ interface has been a promising two-dimensional system with SOC where PHE and AMR can be observed. We envisage many more two dimensional electron systems with large SOC in the years to come where PHE can be observed. 

\acknowledgements
The author thanks Dhavala Suri for useful discussions and DST-INSPIRE Faculty Award (Faculty Reg. No.~:~IFA17-PH190) for financial support. 
\bibliography{ref_phe}

\end{document}